\documentclass[aps,tightenlines,
preprint,groupedaddress,showpacs,nofootinbib]{revtex4}%
\usepackage{graphicx}
\usepackage{amsmath}
\usepackage{bm}

\begin{document}


\title{Relativistic corrections to gluon fragmentation into  
spin-triplet $\bm{S}$-wave quarkonium
}


\author{Geoffrey T. Bodwin and Jungil Lee\footnote{Permanent address: 
Department of Physics, Korea University, Seoul 136-701, Korea.}}
\affiliation{
HEP Division,
Argonne National Laboratory, 9700 South Cass Avenue, Argonne, IL 60439
}


\date{\today}

\begin{abstract}
We use the NRQCD factorization formalism to calculate the relativistic
corrections to the fragmentation function for a gluon fragmenting into a
spin-triplet $S$-wave heavy quarkonium. We make use of the
gauge-invariant formulation of the fragmentation function of Collins and
Soper. The color-octet contribution receives a large, negative
relativistic correction, while the color-singlet contribution receives a
large, positive relativistic correction. The considerable decrease in
the color-octet contribution requires a corresponding increase in the
phenomenological value of the leading color-octet matrix element in
order to maintain a fit to the Fermilab Tevatron data.
\end{abstract}


\maketitle

\section{Introduction\label{intro}}
In the nonrelativistic QCD (NRQCD) factorization approach, the rate of
semi-inclusive quarkonium production at large transverse momentum
($p_T$) is given as a sum of products of short-distance coefficients and
NRQCD matrix elements \cite{BBL}. The short-distance coefficients are
calculable as perturbative series in the strong-coupling constant
$\alpha_s$, while the production matrix elements, at least so far, must
be determined by comparison with experimental data. The field theoretic
ingredients that form the basis for the NRQCD factorization approach for
quarkonium production are the collinear factorization of hard-scattering
processes at large $p_T$ \cite{Collins:1987pm,Collins:gx,drell-yan} and
the decomposition of the factored semi-inclusive production rate into
NRQCD operator matrix elements and short-distance coefficients
\cite{BBL}. The NRQCD factorization formalism predicts that the NRQCD
matrix elements are universal (process independent), and it also leads
to a set of rules \cite{BBL} for the scaling of matrix elements and
interactions with $v$, where $v$ is the heavy-quark or heavy-antiquark
velocity in the quarkonium rest frame. ($v^2\approx 0.3$ for the
$J/\psi$, and $v^2\approx 0.1$ for the $\Upsilon$.) The confrontations
of these expected properties of the NRQCD factorization formulas with
experimental data are among the key tests of the NRQCD approach.

The production of quarkonium at large $p_T$ in $p\bar p$ collisions
provides a particularly important test of the NRQCD factorization
approach. At a $p_T$ of several times the quarkonium mass, the quarkonium
production cross section is dominated by a process in which a gluon
fragments into a quarkonium \cite{BY-gfrag-LO}. Other processes are
suppressed by at least $1/p_T^2$ in the cross section. Furthermore, the
process in which the gluon fragments into the quarkonium through a
color-octet heavy quark-antiquark ($Q\overline Q$) channel dominates
because it is enhanced by a factor $v^4/\alpha_s^2$ relative to the
color-singlet process \cite{BF}. It was pointed out by Cho and Wise
\cite{Cho-Wise} that a quarkonium that is produced through the color-octet
process should have a substantial transverse polarization. The reason
for this transverse polarization is that the fragmenting gluon is nearly
on its mass shell, and, hence, is nearly completely transversely
polarized. In the color-octet fragmentation process, that polarization
is passed on to a produced $Q\overline Q$ pair, which then evolves into a
quarkonium state. According to the velocity-scaling rules of NRQCD
\cite{BBL}, the evolution of the $Q\overline Q$ pair into a quarkonium
state is dominated by non-spin-flip interactions, which preserve the
transverse polarization. Spin-flip interactions are suppressed by at
least $v^2$.

The prediction of substantial quarkonium transverse polarization at
large $p_T$ relies not only on the validity of the NRQCD factorization
formulas for quarkonium production, but also on the universality of the
NRQCD matrix elements and the velocity-scaling rules. Therefore,
experimental measurements of the polarization prediction test many of
the essential features of the NRQCD factorization formalism. Such
measurements are also important in that they can discriminate between
the NRQCD factorization approach and the color-evaporation model, which
predicts zero polarization for the produced quarkonium.

There have been several calculations, based on the NRQCD factorization
approach, of the polarization of the produced quarkonium as a function
of $p_T$. These include calculations of $J/\psi$ polarization at leading
order in $\alpha_s$ \cite{Cho-Wise}, at next-to-leading order in
$\alpha_s$ \cite{beneke-kramer-pol}, and at next-to-leading order in
$\alpha_s$, including the effects of feeddown from $\psi'$ and $\chi_c$
states \cite{BKL-psi-pol}; calculations of $\psi'$ polarization at
next-to-leading order in $\alpha_s$
\cite{beneke-kramer-pol,beneke-rothstein-pol,leibovich-pol,BKL-psi-pol};
and calculations of $\Upsilon$ polarization at next-to-leading order in
$\alpha_s$ \cite{BL-ups-pol}. These calculations are generally in
agreement with each other, although they are subject to large
theoretical uncertainties, which arise mainly from uncertainties in the
color-octet NRQCD matrix elements. The Collider Detector at Fermilab (CDF) 
data for $J/\psi$ and $\psi'$
polarization \cite{CDF-psi-pol} do not support either the prediction of
substantial transverse polarization at large $p_T$ or the prediction of
increasing transverse polarization with increasing $p_T$. However, it
must be said that only the points at the highest $p_T$ are more than
$1.5$ standard deviations away from the predictions in
Ref.~\cite{BKL-psi-pol}. In the case of the $\Upsilon$ polarization, the
CDF data \cite{CDF-ups-pol} agree with the prediction \cite{BL-ups-pol}.
However, the experimental error bars are too large to allow one to draw
firm conclusions about the presence of substantial transverse
polarization.

In light of the possible discrepancy between the predictions for $J/\psi$
and $\psi'$ polarization and the CDF data, it seems worthwhile to
investigate the sizes of uncalculated contributions to the theoretical
cross sections. Among such contributions are relativistic corrections,
which begin in relative order $v^2$. It is known in the case of
quarkonium decay rates that such corrections can be of the same size as
the leading contribution \cite{large-rel}. Investigations of relativistic
corrections to quarkonium decays of order $v^4$ \cite{Bodwin:2002hg}
suggest that the order-$v^2$ contribution gives the bulk of the
correction to the leading contribution.

In this paper, we calculate the relativistic corrections of relative
order $v^2$ to the short-distance coefficients of both the color-octet
and color-singlet terms in the fragmentation function for a gluon to
fragment into a spin-triplet quarkonium state. We carry out the
calculation at leading order in $\alpha_s$. In computing the
short-distance coefficients, we make use of the Collins-Soper definition
of the fragmentation function in terms of a quantum chromodynamic (QCD)
operator matrix element \cite{CS}. In carrying out this calculation, we
confirm the results for the short-distance coefficients at leading
order in $v$ \cite{BY-gfrag-LO}. As a check of our methods, we also
compute the correction of relative order $v^2$ to the three-gluon decay
rate of a spin-triplet quarkonium state and find that it agrees with the
results of Refs.~\cite{KM-ups-3g,Bodwin:2002hg}. We find that the
relativistic correction to the color-octet fragmentation function is
large and negative, while the correction to the color-singlet
fragmentation function is large and positive.

The remainder of this paper is organized as follows. In
Sec.~\ref{sec:fragmentation} we review the Collins-Soper definition of
the fragmentation function, and in Sec.~\ref{sec:factorization} we use
the NRQCD factorization formalism to write the expression for the
fragmentation function in terms of NRQCD operator matrix elements and
perturbatively calculable short-distance coefficients. In
Sec.~\ref{sec:short-distance} we describe the calculation of the
short-distance coefficients. In Secs.~\ref{sec:octet} and
\ref{sec:singlet} we compute, respectively, the short-distance
coefficients for the color-octet and color-singlet parts of the
fragmentation function for a gluon fragmenting into a ${}^3S_1$
quarkonium state, through relative order $v^2$. Finally, we discuss the
implications of our results in Sec.~\ref{sec:discuss}.
\section{Collins-Soper definition of the fragmentation function 
\label{sec:fragmentation}}
In this section we make use of the Collins-Soper definition of the
fragmentation function \cite{CS} to write, in terms of NRQCD operator
matrix elements and short-distance coefficients, the fragmentation
function for a gluon to fragment into a quarkonium state. The
Collins-Soper definition was first used in a calculation of a quarkonium
fragmentation function by Ma \cite{ma}.

Here, and throughout this paper, we use the following light-cone 
coordinates for a four-vector $V$: 
\begin{subequations}
\begin{eqnarray}
V&=&(V^+,V^-,\bm{V}_\bot)=(V^+,V^-,V^1,V^2),\\
V^+&=&(V^0+V^3)/\sqrt{2},\\
V^-&=&(V^0-V^3)/\sqrt{2}.
\end{eqnarray}
\end{subequations}
The scalar product of two four-vectors $V$ and $W$ is then 
\begin{equation}
V\cdot W=V^+W^-+V^-W^+-\bm{V}_\bot\cdot\bm{W}_\bot.
\end{equation}

The fragmentation function $D_{g \to H}(z,\mu)$ is the probability for a
gluon that has been produced in a hard-scattering process to decay into
a hadron $H$ carrying a fraction $z$ of the $+$ component of the gluon's
momentum. This function can be defined, in a light-cone gauge, in terms
of the matrix element of a bi-local operator involving two gluon field
strengths \cite{CFP}. In Ref.~\cite{CS}, Collins and Soper introduced a
gauge-invariant definition of the gluon fragmentation function:
\begin{eqnarray}
D_{g \to H}(z,\mu) &=&
\frac{-g_{\mu \nu}z^{d-3} }{ 2\pi k^+ (N_c^2-1)(d-2) }
\int_{-\infty}^{+\infty} dx^- e^{-i k^+ x^-}
\nonumber\\
&&\times
\langle 0 | G^{+\mu}_c(0)
\mathcal{E}^\dagger(0^-)_{cb} \; \mathcal{P}_{H(z k^+,\bm{0}_\perp)} \;
\mathcal{E}(x^-)_{ba} G^{+ \nu}_a(0^+,x^-,\bm{0}_\perp) | 0 \rangle\;.
\label{eq:D-def}
\end{eqnarray}
Here, $G_{\mu\nu}$ is the gluon field-strength tensor, $k$ is the
momentum of the field-strength tensor, and $d=4-2\epsilon$ is the number
of space-time dimensions. There is an implicit average over the color and
polarization states of the initial gluon. The parameter $\mu$ is the
factorization scale, which appears implicitly through the dimensional
regularization of the operator matrix element. The operator
$\mathcal{P}_{H(P^+,\bm{P}_\perp)}$ is a projection onto states
that, in the asymptotic future, contain a hadron $H$ with momentum $P =
\big(P^+,P^-=(M^2+ P_\perp^2)/(2P^+),\bm{P}_\perp\big)$, where $M$ is the mass
of the hadron:
\begin{equation}
\mathcal{P}_{H(P^+,\bm{0}_\perp)}=
\sum_{X} |H(P^+,\bm{0}_\perp)+X\rangle \langle H(P^+,\bm{0}_\perp)+X|.
\label{eq:PH}
\end{equation}
The eikonal operator $\mathcal{E}(x^-)$ is a path-ordered exponential of
the gluon field that makes the expression (\ref{eq:D-def}) gauge
invariant by connecting the different space-time positions of the gluon
field strengths:
\begin{equation}
\mathcal{E}(x^-)_{ba} \;=\; \textrm{P} \exp
\left[ +i g \int_{x^-}^\infty dz^- A^+(0^+,z^-,0_\perp) \right]_{ba},
\label{eq:E}
\end{equation}
where $g$ is the QCD coupling constant, and $A^\mu(x)$ is the gluon
field. Both $A_\mu$ and $G_{\mu\nu}$ are SU(3)-matrix valued, with the
matrices in the adjoint representation. The manifest gauge invariance of
expression (\ref{eq:D-def}) allows us to make use of the Feynman
gauge in order to simplify the calculation.

The definition (\ref{eq:D-def}) is invariant under boosts along the
longitudinal direction \cite{CS}. The fragmentation function is defined
in a frame in which the transverse momentum of the hadron $H$ is
vanishes: $\bm{P}_\perp=\bm{0}_\perp$. However, the
fragmentation function is also invariant under boosts that change the
transverse momentum of the hadron, while leaving the $+$ components of
all momenta unchanged \cite{CS}. [This can be seen from the fact that
the definition (\ref{eq:D-def}) is manifestly covariant, except for
dependences on the $+$ components of some quantities and on the dummy
variable $x^-$.] An explicit construction of such a Lorentz
transformation is given in Appendix~\ref{app:boost}. However, for the
purposes of this calculation, we find it convenient to work in the frame
in which $\bm{P}_\perp=\bm{0}_\perp$.

In general, the fragmentation function (\ref{eq:D-def}) involves the
long-distance dynamics of the evolution of gluon into a quarkonium state
and, hence, is a nonperturbative quantity. However, we may evaluate the
short-distance part of evolution of the gluon into a $Q\overline Q$ state
$H$ as a power series in $\alpha_s$. A convenient set of Feynman rules
for the perturbative expansion of Eq.~(\ref{eq:D-def}) is given in
Ref.~\cite{CS}. For the purposes of this calculation, we need only the
standard QCD Feynman rules and the special rule for the creation of a
gluon by the operator $G_a^{+\nu}$ in Eq.~(\ref{eq:D-def}). The latter
rule, in momentum space, is a factor
\begin{equation}
+ik^+
\left(g^{\nu\alpha}-\frac{Q^\nu n^\alpha}{k^+}\right)
\delta_{ab},
\label{eq:gL}
\end{equation}
where $k$ is the momentum of the field-strength tensor, $Q$ is the
momentum of the gluon, $\alpha$ and $a$ are the vector and color
indices, respectively, of the created gluon, and $b$ is the color index
of the eikonal line. In the absence of interactions with the eikonal
line, $k=Q$.

In the definition (\ref{eq:D-def}), the transverse momentum of the
hadron $H$ is fixed. Therefore, the phase space is the product of the
phase-space factors of all of the other final-state particles and the 
light-cone-energy-conserving delta function that is implied by the
integration over $x^-$ in Eq.~(\ref{eq:D-def}):
\begin{equation}
d\Phi_n=\frac{4\pi M }{S}\;\delta\left(k^+-P^+-\sum_{i=1}^{n}a_i^+\right)
\prod_{i=1}^n\frac{da_i^+ d^{d-2}\bm{a}_\perp}{2a^+(2\pi)^{d-1}},
\label{eq:phase-space-n}
\end{equation}
where $S$ is the statistical factor for identical particles in the final
state, $a_i$ is the momentum of the $i$th final-state particle, and the
product is over all of the final-state particles except $H$. Since we use
nonrelativistic normalization for the hadron $H$, a factor $2M$ has been
included in the phase space in order to cancel the relativistic
normalization of $H$ in the definition (\ref{eq:D-def}). In the
remainder of this paper, we use nonrelativistic normalization for the
hadron $H$ and heavy quark $Q$ and antiquark $\overline Q$, and we use
relativistic normalization for all of the other particles.
\section{NRQCD Factorization\label{sec:factorization}}

The fragmentation of a gluon into a heavy-quarkonium state $H$ involves
many momentum scales, ranging from the factorization scale of the
fragmentation function $\mu$, which we assume to be of the order of the
heavy-quark mass $m$ or greater, to momenta much smaller than $m$, for
which nonperturbative effects are large. The NRQCD factorization
formalism allows one to make a systematic separation of momentum scales
of order $m$ and larger from scales of order $m v$ or smaller. Following
this approach \cite{BBL}, we write the fragmentation function of a gluon
fragmenting into a heavy quarkonium in the form
\begin{equation}
D_{g \to H}(z,\mu)=
\sum_{n} 
\left[
d_{n}(z,\mu) \langle \mathcal{O}_{n}^H \rangle\;
+
d'_{n}(z,\mu) \langle \mathcal{P}_{n}^H \rangle\;
\right]+O(v^3),
\label{eq:NRQCD-fact-H}
\end{equation}
where the $\mathcal{O}_{n}^H$ and $\mathcal{P}_{n}^H$ are NRQCD
operators, and the $d_{n}(z,\mu)$ and $d'_{n}(z,\mu)$ are short-distance
coefficients. The first term on the right-hand side of
Eq.~(\ref{eq:NRQCD-fact-H}) contains the operator matrix elements of
leading order in $v$, and the second term contains the operator matrix
elements of relative order $v^2$. The index $n$ represents the quantum
numbers of the operator. The matrix elements $\langle \mathcal{O}_{n}^H
\rangle$ and $\langle \mathcal{P}_{n}^H \rangle$ are defined in the rest
frame of $H$ and are given, in the case of a ${}^3S_1$ state $H$, by the
vacuum expectation values of the following four-quark operators:
\begin{subequations}
\label{eq:NRQCD-ME}
\begin{eqnarray}
\mathcal{O}_{1}^H  &=&
\chi^\dagger \sigma^i\psi \; \sum_{X} |H+X\rangle \langle H+X| \;
        \psi^\dagger \sigma^i \chi ,
\\
\mathcal{O}_{8}^H &=&
\chi^\dagger \sigma^i T^a\psi \;  
\sum_{X} |H+X\rangle \langle H+X|\;
        \psi^\dagger \sigma^i T^a\chi ,
\\
\mathcal{P}_{1}^H &=&
\frac{1}{2m^2}\left[\chi^\dagger \sigma^i 
\left({\textstyle\frac{i}{2}}\tensor{\bm{D}}\right)^2\psi \; 
\sum_{X} |H+X\rangle \langle H+X| \;
        \psi^\dagger \sigma^i \chi +\textrm{H.~c.}\right]\;,
\\
\mathcal{P}_{8}^H &=&
\frac{1}{2m^2}\left[\chi^\dagger \sigma^i T^a
\left({\textstyle\frac{i}{2}}\tensor{\bm{D}}\right)^2\psi \;
\sum_{X} |H+X\rangle \langle H+X| \;
        \psi^\dagger \sigma^i T^a\chi +\textrm{H.~c.}\right],
\end{eqnarray}
\end{subequations}
where $m$ is the heavy-quark mass, $\psi$ is the Pauli field that
annihilates a $Q$, $\chi$ is a Pauli field that creates a $\overline{Q}$, and
$\chi^\dagger\tensor{\bm{D}}\psi=\chi^\dagger(\bm{D}\psi)
-(\bm{D}\chi)^\dagger\psi$. The sum is over all final states that
contain the specified quarkonium state $H$. The NRQCD matrix
elements are nonperturbative in nature, but they are universal, in that
the same matrix elements describe inclusive production of ${}^3S_1$
quarkonium states in other high-energy processes.

\section{Short-distance coefficients \label{sec:short-distance}}

The short-distance coefficients in Eq.~(\ref{eq:NRQCD-fact-H}) are
independent of the long-distance dynamics of hadronization and,
consequently, they are independent of the hadronic state $H$. Therefore,
they can be calculated by examining the fragmentation function for the
case in which the state $H$ is a free $Q\overline Q$ state. We take
the momenta of the $Q$ and $\overline Q$ to be $p=P/2+q$ and
$\bar{p}=P/2-q$, respectively. The heavy quark has three-momentum $\bm{q}$
in the $Q \overline{Q}$ rest frame, and, so, the invariant mass of the
$Q\overline Q$ state is $P^2 = 4E^2$, where $E=\sqrt{m^2+\bm{q}^2}$.
Since the momentum $q$ is fixed, the phase space of the $Q \overline{Q}$
pair is that of a single particle with momentum $P$, and it is omitted in
the phase-space integrations in the computation of the fragmentation
function. The fragmentation function for this free $Q\overline Q$ state
is
\begin{equation}
D_{g \to Q\overline{Q}}(z,\mu) =
\sum_{n} 
\left[
d_{n}(z,\mu) \langle \mathcal{O}_{n}^{Q\overline{Q}} \rangle
+
d'_{n}(z,\mu) \langle \mathcal{P}_{n}^{Q\overline{Q}} \rangle
\right]+O(v^3)\;.
\label{eq:NRQCD-fact-QQ}
\end{equation}
The definitions of the NRQCD matrix elements $\langle
\mathcal{O}_{n}^{Q\overline{Q}} \rangle$ and $\langle
\mathcal{P}_{n}^{Q\overline{Q}} \rangle$ are same as those in
Eq.~(\ref{eq:NRQCD-ME}), except for the replacement $H\to 
Q\overline{Q}$. The short-distance coefficients $d_{n}(z,\mu)$
and $d'_{n}(z,\mu)$, which are common to both
Eq.~(\ref{eq:NRQCD-fact-H}) and Eq.~(\ref{eq:NRQCD-fact-QQ}), can be
obtained by comparing a perturbative calculation of the fragmentation
function on the left-hand side of Eq.~(\ref{eq:NRQCD-fact-QQ}) with a
perturbative calculation of the NRQCD matrix elements on the right-hand
side of Eq.~(\ref{eq:NRQCD-fact-QQ}). While the fragmentation function
and the matrix elements in Eq.~(\ref{eq:NRQCD-ME}) may both display
sensitivity to long-range (infrared) interactions, that sensitivity
cancels in the short-distance coefficients.

In determining the coefficients of $\langle
\mathcal{O}_{n}^{Q\overline{Q}} \rangle$ and $\langle
\mathcal{P}_{n}^{Q\overline{Q}} \rangle$ in $D_{g \to
Q\overline{Q}}(z,\mu)$, it is convenient to make use of projection
operators for the spin and color states of the $Q\overline{Q}$ pair. The
projection operators for the $Q\overline{Q}$ pair in the color-singlet
and color-octet configurations are
\begin{subequations}
\label{eq:PJ-color}
\begin{eqnarray}
\Lambda_1&=&\frac{1}{\sqrt{N_c}} 
\delta_{ji},\label{eq:PJ-color-singlet}\\
\Lambda^a_8&=&\sqrt{2} T^a_{ji},
\label{eq:PJ-color-octet}
\end{eqnarray}
\end{subequations}
where $i$ and $j$ are the quark color indices in the fundamental
(triplet) representation, $a$ is the octet color index, and $N_c=3$.
Spin-projection operators, accurate to all orders in $v$, have been
given in Ref.~\cite{Bodwin:2002hg}.\footnote{Projection operators that 
are accurate to lowest order in the nonrelativistic expansion have been
given previously \cite{proj-old}.} For the spin-triplet state, the
projection operator is
\begin{equation}
\Lambda(P,q,\epsilon^*)=\Lambda^\alpha(P,q)\epsilon^*_\alpha=N
(\not\!{\overline{p}}-m)\not\!{\epsilon}^*\frac{\not\!{P}+2E}{4E}
(\not\!{p}+m),
\label{eq:PJ-spin}
\end{equation}
where $\epsilon$ is the polarization of the heavy-quark pair, and
$N=[2\sqrt{2}E(E+m)]^{-1}$. Note that we use nonrelativistic
normalization for the heavy-quark spinors. If $\mathcal{C}$ is the full
QCD amplitude for a process, then the spin-triplet part of that
amplitude is $\mathcal{M}=\textrm{Tr}(\mathcal{C}\Lambda)$. 

The $S$-wave part of $\mathcal{M}$ is
\begin{equation}
\mathcal{M}_{S\textrm{-wave}}=
\mathcal{M}_0
+\frac{\bm{q}^2}{m^2}\mathcal{M}_2+O(\bm{q}^4),
\label{eq:M-S-wave}
\end{equation}
where the first two terms on the right-hand side of Eq.~(\ref{eq:M-S-wave}) 
are the leading and first subleading terms in the $v$ expansion.
Here, 
\begin{subequations}
\label{eq:M-S-2}
\begin{eqnarray}
\mathcal{M}_0&=&\left.\mathcal{M}\right|_{\bm{q}\to \bm{0}}\;,\\
\mathcal{M}_2&=&\left.
             \frac{m^2I^{\alpha\beta} }{2(d-1)}
                           \frac{\partial^2 \mathcal{M}}
                                {\partial q^\alpha\partial q^\beta}
\right|_{\bm{q}\to \bm{0}}\;,
\end{eqnarray}
\end{subequations}
where 
\begin{equation}
I^{\alpha\beta}=-g^{\alpha\beta}+P^\alpha P^\beta/(4E^2).
\label{I-ab}
\end{equation}
The matrix elements of the $Q\overline Q$ NRQCD operators are normalized as
\begin{subequations}
\label{eq:normalization}
\begin{eqnarray}
\langle \mathcal{O}_1^{Q\overline{Q}}\rangle
&=&2(d-1)N_c,
\label{eq:normalization-a}
\\
\langle \mathcal{O}_8^{Q\overline{Q}}\rangle
&=&(d-1)(N_c^2-1),
\label{eq:normalization-b}
\\
\langle \mathcal{P}_n^{Q\overline{Q}}\rangle&=&\frac{\bm{q}^2}{m^2}
\langle \mathcal{O}_n^{Q\overline{Q}}\rangle.
\label{eq:normalization-c}
\end{eqnarray}
\end{subequations}
The quantities that we calculate in this paper are all finite. Thus, we
may put $d=4$ in Eqs.~(\ref{eq:D-def}), (\ref{eq:phase-space-n}), and
(\ref{eq:normalization}).
\section{Color-octet contribution\label{sec:octet}}
\begin{figure}
\includegraphics[height=5cm]{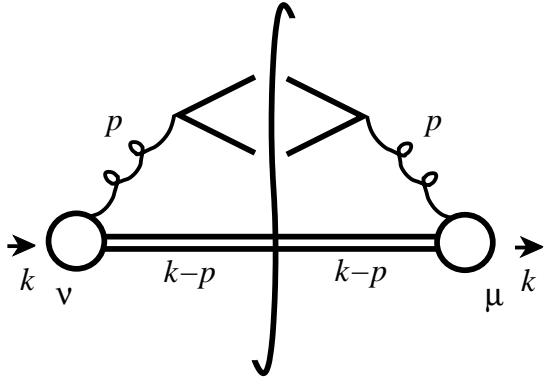}
\caption{Feynman diagram for the color-octet contribution at leading
order in $\alpha_s$ to the fragmentation function for a gluon
fragmenting into a color-octet spin-triplet $Q\overline{Q}$ pair.
\label{fig:g-QQ}}
\end{figure}

Let us calculate the color-octet contribution to the fragmentation of a
gluon into a $Q\overline Q$ pair in full QCD. The Feynman diagram for the
color-octet part of the fragmentation function in leading order in
$\alpha_s$ is shown in Fig.~\ref{fig:g-QQ}. The circles represent the
the gluon field strengths and the double lines represent the eikonal
operator. The momentum $k = (k^+,k^-,\bm{k}_\perp)$ flows into the
circle on the left and out the circle on the right. The vertical line
represents the final-state cut. In leading order in $\alpha_s$, the
final state consists of a $Q \overline{Q}$ pair with total momentum $P =
(z k^+, P^2/(2z k^+), \bm{0}_\perp)$.
The Feynman diagram in Fig.~\ref{fig:g-QQ} includes 
relativistic corrections to all orders in $v$.
The phase-space element is obtained by setting $n=0$ and $S=1$
in Eq.~(\ref{eq:phase-space-n}): 
\begin{equation}
d\Phi_0=\frac{4\pi M}{k^+}\delta(1-z).
\label{eq:phase-space-0}
\end{equation}
The color-octet contribution can be extracted by tracing over the
projection operator $\Lambda_8^a$ [Eq.~(\ref{eq:PJ-color-octet})] on both
the left-hand and right-hand sides of the cut. Summing over the color
index $a$, we obtain the color factor $(N_c^2-1)/2$. The spin-triplet
contribution can be extracted by tracing over the projection operator
$\Lambda^\alpha$ [Eq.~(\ref{eq:PJ-spin})] on the left-hand side of the
cut and the projection operator $\Lambda^\beta$ on the right-hand side of
the cut. Multiplying by the prefactor in Eq.~(\ref{eq:D-def}) and the
phase-space element (\ref{eq:phase-space-0}), we obtain
\begin{equation}
D_8^{Q\overline{Q}}(z)=
-\frac{\pi \alpha_s M \mathcal{M}^{\alpha\sigma}_{S\textrm{-wave}} 
                      \mathcal{M}^{\beta\tau *}_{S\textrm{-wave}} }
        {8E^4}
g^{\mu\nu}I_{\sigma\tau}
\left(g_{\nu\alpha}-\frac{P_\nu n_\alpha}{k^+}\right)
\left(g_{\mu\beta}-\frac{P_\mu n_\beta }{k^+}\right)
\delta(1-z),
\label{D-octet}
\end{equation}
where $I_{\sigma\tau}$ is given in Eq.~(\ref{I-ab}), and 
\begin{equation}
\mathcal{M}^{\alpha\sigma}_{S\textrm{-wave}}=
\textrm{Tr}\left(\gamma^\alpha \Lambda^\sigma\right)_{S\textrm{-wave}}
=
-\sqrt{2} g^{\alpha\sigma}
\left(1-\frac{\bm{q}^2}{6m^2}\right)+O(v^4).
\label{octet-spin-prog}
\end{equation}
In the last equality in Eq.~(\ref{octet-spin-prog}), we have made use of
Eq.~(\ref{eq:M-S-2}). Contracting vector indices, factoring out the
normalization of the octet matrix elements given in
Eqs.~(\ref{eq:normalization-b}) and (\ref{eq:normalization-c}), and
setting $M=2E$, we obtain
\begin{eqnarray}
D_8^{Q\overline{Q}}(z)&=&
\frac{\pi\alpha_s \delta(1-z)}
        {3(N_c^2-1) m^3}
\left[\langle \mathcal{O}^{Q\overline{Q}}_8\rangle
-\frac{11}{6} \langle \mathcal{P}^{Q\overline{Q}}_8\rangle\right]
+O(v^4)\nonumber\\
&=&d_8(z)\,\langle \mathcal{O}^{Q\overline{Q}}_8\rangle                
+d_8'(z)\,\langle \mathcal{P}^{Q\overline{Q}}_8\rangle+O(v^4),
\label{eq:d8-ans}
\end{eqnarray}
where $d_8(z)$ and $d_8'(z)$ are the short-distance coefficients. We
obtain the contribution to the fragmentation function for a gluon
fragmenting into a quarkonium $H$ by replacing the $Q\overline Q$
operator matrix elements in Eq.~(\ref{eq:d8-ans}) by $H$ operator matrix
elements:
\begin{equation}
D_8^H(z)=
\frac{\pi\alpha_s \delta(1-z)}
        {3(N_c^2-1) m^3} \langle \mathcal{O}^{H}_8\rangle
\left[1-\frac{11}{6} v_8^2+O(v^4)\right],
\label{color-octet-final}
\end{equation}
where 
\begin{equation}
v_8^2=\langle \mathcal{P}^{H}_8\rangle /\langle
\mathcal{O}^{H}_8\rangle.
\label{v-sq-octet}
\end{equation}
Our result for the term of leading order in
$v$ in Eq.~(\ref{color-octet-final}) is in agreement with the results of
Refs.~\cite{Braaten:1994kd,BL:gfrag-NLO}.
\section{Color-singlet contribution\label{sec:singlet}}
\begin{figure}
\includegraphics[height=5cm]{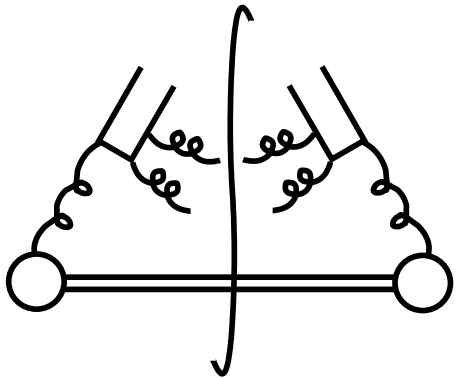}
\caption{One of the Feynman diagrams for the color-singlet contribution
at leading order in $\alpha_s$ to the fragmentation function for a gluon
fragmenting into a color-singlet spin-triplet $Q\overline{Q}$ pair. The
other diagrams are obtained by permuting the connections of the
gluons to the heavy-quark lines. \label{fig:g-QQgg}}
\end{figure}

Now let us calculate the color-singlet part of the fragmentation
function for a gluon fragmenting into a $Q\overline Q$ pair in full QCD.
The Feynman diagram for this process at leading order in $\alpha_s$ is
shown in Fig.~\ref{fig:g-QQgg}. We assign the momenta and 
polarization indices for the particles in this process as follows:
\begin{equation}
g^*(k,\alpha)\to Q\overline{Q}(P,\sigma)
                   + g(a,{\mu_a})+g(b,{\mu_b}),
\label{eq:singlet-process}
\end{equation}
where $k$, $P$, $a$, and $b$ are momenta and $\alpha$, $\sigma$,
$\mu_a$, and $\mu_b$ are polarization indices. Owing to the
color-singlet and spin-triplet quantum number of the $Q\overline Q$
pair, the final state must contain at least two gluons that couple to
the heavy-quark line. Hence, in leading order in $\alpha_s$, there are
no diagrams for this process, in any gauge, in which a gluon couples to
the eikonal line.

In the frame in which Eq.~(\ref{eq:D-def}) is defined,
we can choose the transverse direction so that
\begin{subequations}
\label{eq:momentum-2g}
\begin{eqnarray}
P&=&\left(zk^+,\frac{(2E)^2}{2zk^+},0,0\right),
\\
a&=&\left(yk^+,\frac{a_\perp^2}{2yk^+},a_\perp,0\right),
\\
b&=&\left(wk^+,\frac{b_\perp^2}{2wk^+}
                 ,b_\perp\cos\phi
                 ,b_\perp\sin\phi\right)\;,
\end{eqnarray}
\end{subequations}
where $\phi$ is the azimuthal angle of $b$ relative to $a$. The
variables $y$ and  $w$ are the light-cone fractions of final-state
gluons: $y=a^+/k^+$, and  $w=b^+/k^+$, with $z+y+w=1$. The phase-space
element of the process can be obtained by making the substitutions
$n=2$, $S=2$, $a_1=a$, and $a_2=b$ in Eq. (\ref{eq:phase-space-n}). It
is useful in carrying out the phase-space integration to express the
variables $a_\perp$, $b_\perp$, and $\phi$ in terms of Lorentz-invariant
dimensionless variables $e_a$, $e_b$, and $x$:
\begin{subequations}
\label{ea-eb-x}
\begin{eqnarray}
e_a&=&\frac{P\cdot a}{P^2}=\frac{1}{2}\left(
                    \frac{za_\perp^2}{(2E)^2y} +\frac{y}{z}
                    \right),
\\
e_b&=&\frac{P\cdot b}{P^2}=\frac{1}{2}\left(
                    \frac{zb_\perp^2}{(2E)^2w} +\frac{w}{z}
                    \right),
\\
x&=&\frac{a\cdot b}{P^2}
=\frac{1}{2(2E)^2}\left(
                    \frac{wa_\perp^2}{y} +
                    \frac{yb_\perp^2}{w}
-2a_\perp b_\perp\cos\phi
                    \right)\;.\label{x}
\end{eqnarray}
\end{subequations}
Note that only the invariant variable $x$ depends on the angle $\phi$.
In terms of the variables $e_a$ and $e_b$, the phase space is
\begin{subequations}
\label{eq:phase-space-2}
\begin{equation}
d\Phi_2=
\frac{M (2E)^4}{4z^2k^+(2\pi)^3}\int_\Phi\;,
\label{eq:phase-space-20}
\end{equation}
where
\begin{equation}
\int_\Phi\equiv
\int_{0}^{\infty} de_a\;
\int_{0}^{\infty} de_b\;
\int_{0}^{1-z}dy
\int_0^{2\pi} \frac{d\phi}{2\pi}
\;
\theta\left( e_a- \frac{y}{2z}\right)
\theta\left( e_b- \frac{w}{2z}\right)
\label{eq:phi-integral}
\end{equation}
\end{subequations}
and we have integrated over the azimuthal angle of $a$ and the variable
$w$. The $\theta$ functions in Eq.~(\ref{eq:phi-integral}) impose the
requirement of the positivity of the variables $a_\perp^2$ and
$b_\perp^2$. We defer the discussion of the change of variables from
$\phi$ to $x$ to Appendix~\ref{phase-space}.

The color-singlet contribution can be extracted by tracing over the
projection operator $\Lambda_1$  [Eq.~(\ref{eq:PJ-color-singlet})] on
both sides of the final state cut. Using charge-conjugation symmetry to
relate diagrams involving permutations of the gluon connections to the
heavy-quark lines (Fig.~\ref{fig:g-QQgg}), we find that the color factor 
is
\begin{equation}
\left(\frac{d^{abc}}{4\sqrt{N_c}}\right)^2
=\frac{(N_c^2-4)(N_c^2-1)}{16N^2_c}\;.
\end{equation}
The spin-triplet contribution can be extracted by tracing over the
projection operator $\Lambda^\alpha$ [Eq.~(\ref{eq:PJ-spin})] on the
left-hand side of the cut and the projection operator $\Lambda^\beta$ on
the right-hand side of the cut. Multiplying by the prefactor in
Eq.~(\ref{eq:D-def}) and the phase-space element
(\ref{eq:phase-space-20}), we obtain
\begin{equation}
D_1^{Q\overline{Q}}(z)=
\frac{(N_c^2-4)\alpha_s^3 M}{32\pi z^{5-d} N_c^2}
\int_\Phi
\frac{\bm{\Xi}\cdot\mathcal{A}}{(1+2e_a+2e_b+2x)^2}\;,
\label{eq:dz_integral}
\end{equation}
where the scalar product in $\bm{\Xi}\cdot\mathcal{A}$ denotes the sum
over all the repeated vector indices appearing in the product of the
tensors $\bm{\Xi}$ and $\mathcal{A}$. These tensors are defined as
\begin{subequations}
\label{full-QCD-integrand}
\begin{eqnarray}
\bm{\Xi}&=&
-g_{\mu\nu}\;
g_{\mu_a\nu_a}\;
g_{\mu_b\nu_b}\;
I_{\sigma\tau}
\left(g_{\nu\alpha}-\frac{k_\nu n_{\alpha}}{k^+}\right)
\left(g_{\mu\beta}-\frac{k_\mu n_{\beta}}{k^+}\right)\;, 
\\
\mathcal{A}&=&
\mathcal{M}\otimes \mathcal{M}^*\equiv
\mathcal{M}^{\alpha \mu_a \mu_b\sigma}_{S\textrm{-wave}}
\;
\mathcal{M}^{*\beta \nu_a \nu_b\tau}_{S\textrm{-wave}}\;,
\label{eq:S-wave}
\\
\mathcal{M}^{\alpha \mu_a \mu_b\sigma}&=&
\textrm{Tr}\left[
\gamma^{\alpha}
\frac{1}{\not\!{p}-\not\!{k}-m}
\gamma^{\mu_a}
\frac{1}{\not\!{\overline{p}}-\not\!{b}-m}
\gamma^{\mu_b}\Lambda^\sigma(P,q)
\right]+\;5\;\textrm{perm.}\;,
\label{eq:perm}
\end{eqnarray}
\end{subequations}
where $I_{\sigma\tau}$ is given in Eq.~(\ref{I-ab}), and the terms
labeled ``perm.''\ in Eq.~(\ref{eq:perm}) are generated by permuting
the gluon momenta and polarization indices $(-k,\alpha)$, $(a,\mu_a)$,
$(b,\mu_b)$. In Eq.~(\ref{full-QCD-integrand}), we have made use of the
gauge invariance of the Collins-Soper form of the fragmentation function
(\ref{eq:D-def}) to carry out the calculation in the Feynman gauge.
We obtain the $S$-wave contributions of leading order in $v$ and 
of relative order $v^2$ in Eq.~(\ref{eq:S-wave}) by applying
Eqs.~(\ref{eq:M-S-wave}) and (\ref{eq:M-S-2}) to the spin-triplet
amputated amplitude (\ref{eq:perm}). 

Substituting $N_c=3$, using Eq.~(\ref{eq:normalization}) to factor out
the NRQCD matrix element $\langle
\mathcal{O}_1^{Q\overline{Q}}\rangle=6N_c$, and retaining terms through
relative order $v^2$ in $\bm{\Xi}\cdot\mathcal{A}$ and in the phase-space
factor $M=2E$, we obtain
\begin{eqnarray}
D_1^{Q\overline{Q}}(z)&=&
\frac{5\alpha_s^3 m}{2592\pi z}
\int_\Phi
\frac{F_0\,\langle \mathcal{O}_1^{Q\overline{Q}}\rangle
+\left(\frac{1}{2}F_0+F_2\right)\langle
\mathcal{P}_1^{Q\overline{Q}}\rangle}
{(1+2e_a+2e_b+2x)^2}+O(v^4)\nonumber\\
&=&d_1(z)\,\langle \mathcal{O}_1^{Q\overline{Q}}\rangle
+d_1'(z)\,\langle\mathcal{P}_1^{Q\overline{Q}}\rangle+O(v^4),
\label{singlet-frag-QQ}
\end{eqnarray}
where $d_1(z)$ and $d_1'(z)$ are the short-distance coefficients, and
\begin{subequations}
\label{eq:integrand}
\begin{eqnarray}
F_0&=&
\left.
\bm{\Xi}\cdot \left(\mathcal{M}_0 \otimes 
\mathcal{M}_0^*\right)
\right|_{\bm{q}\to \bm{0}}\;,
\label{eq:integrand-F0}
\\
F_2&=&m^2
\frac{\partial}{\partial \bm{q}^2}
\left.\left\{
\bm{\Xi}\cdot \textrm{Re}\left[\left(\mathcal{M}_0
                +2\frac{\bm{q}^2}{m^2}\mathcal{M}_2\right)\otimes
                             \mathcal{M}_0^*\right]\right\}
\right|_{\bm{q}\to \bm{0}}\;.
\label{eq:integrand-F2}
\end{eqnarray}
\end{subequations}
[Note that, although $\mathcal{M}_0$ and $\mathcal{M}_2$ are independent
of $\bm{q}$, the factors $\bm{\Xi}$ in Eq.~(\ref{eq:integrand}) introduce
a dependence on $\bm{q}^2$.] We obtain the contribution to the
fragmentation function for a gluon fragmenting into a quarkonium $H$ by
replacing the $Q\overline Q$ operator matrix elements in
Eq.~(\ref{singlet-frag-QQ}) by $H$ operator matrix elements:
\begin{equation}
D_1^H(z)=
\frac{5\alpha_s^3 m \langle \mathcal{O}_1^H\rangle}
{2592\pi z}
\int_\Phi
\frac{F_0
+\left(\frac{1}{2}F_0+F_2\right)v_1^2}
{(1+2e_a+2e_b+2x)^2}+O(v^4),
\label{singlet-frag-H}
\end{equation}
where $v_1^2$ is given by
\begin{equation}
v_1^2=\langle \mathcal{P}_1^H\rangle/\langle \mathcal{O}_1^H\rangle,
\end{equation}
rather than by Eq.~(\ref{v-sq-octet}).

The evaluations of $F_0$ and $F_2$ in Eqs.~(\ref{eq:integrand}) are
straightforward, but quite involved.  We compute $F_0$ and $F_2$ using
\textsc{reduce} \cite{REDUCE}. As a check, we carry out independent
calculations using the \textsc{feyncalc} package \cite{Mertig:an} in
\textsc{mathematica} \cite{MATH}. By making the replacements
$\bm{\Xi}\to -g_{\mu_a\nu_a}g_{\mu_b\nu_b} g_{\alpha\beta}
I_{\sigma\tau}$ and $k^2\to 0$ and multiplying by the appropriate
factor, we also obtain the decay rate of the process $\Upsilon\to ggg$,
including the relativistic correction. Our results agree with those in
Refs.~\cite{KM-ups-3g,Bodwin:2002hg}. The rather lengthy expressions for
$F_0$ and $F_2$ are given in a \textsc{mathematica} notebook 
file~\cite{nb:integrands}.

\begin{figure}
\includegraphics[height=8cm]{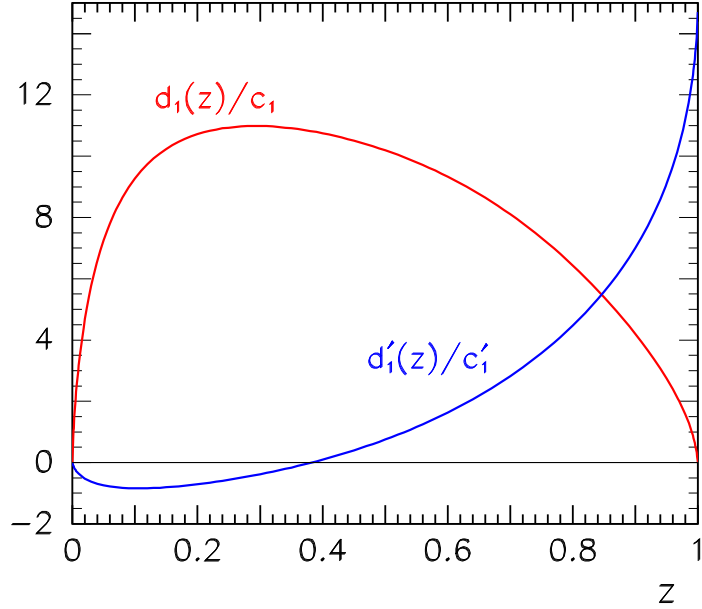}
\caption{The color-singlet short-distance coefficients $d_1(z)$ and
$d_1'(z)$, which are defined in Eq.~(\ref{singlet-frag-QQ}). The scaling
factors in this figure are $c_1=10^{-4}\times\alpha_s^3/m^3$ and 
$c_1'=10^{-3}\times\alpha_s^3/m^3$.
\label{fig:frag-fn}}
\end{figure}

In order to evaluate the integrals in Eq.~(\ref{singlet-frag-H}), we
use Eq.~(\ref{x}) to replace $\phi$ with $x$, as we have already mentioned,
and we carry out the $y$ integration analytically. This
procedure is described in detail in Appendix~\ref{phase-space}. The
integration methods that we use are similar to those in
Ref.~\cite{BY-gfrag-LO}. However, in Ref.~\cite{BY-gfrag-LO}, the
short-distance coefficient $d_1(z)$ is expressed as a two-dimensional
integral, which is then evaluated numerically. In the present
calculation, we carry out the integrations over the three variables
$e_a$, $e_b$, and $x$ numerically. This procedure allows one to
extract information about the energy spectrum of the radiated gluons. 
The results of this numerical integration are presented in 
Fig.~\ref{fig:frag-fn} and in Table~\ref{tab:dz}.
\begin{table}[t]
\caption{\label{tab:dz}%
Numerical values of the color-singlet short-distance coefficients
$d_1(z)$ and $d_1'(z)$, which are defined in Eq.~(\ref{singlet-frag-QQ}).
The scaling factors in this table are $c_1=10^{-4}\times
\alpha_s^3/m^3$ and $c_1'=10^{-3}\times\alpha_s^3/m^3$.
}
\begin{ruledtabular}
\begin{tabular}{lll}
 $z$ & $d_1(z)/c_1$ & $d_1'(z)/c_1'$\\
\hline
0    & \, 0     &  0      \\
0.05 & \, 7.25  &  \!\!\!\!\!$-$0.746  \\
0.10 & \, 9.27  &  \!\!\!\!\!$-$0.842  \\
0.15 &   10.2   &  \!\!\!\!\!$-$0.808  \\
0.20 &   10.7   &  \!\!\!\!\!$-$0.709  \\
0.25 &   10.9   &  \!\!\!\!\!$-$0.564  \\
0.30 &   11.0   &  \!\!\!\!\!$-$0.382  \\
0.35 &   10.9   &  \!\!\!\!\!$-$0.162  \\
0.40 &   10.7   &   0.0974 \\
0.45 &   10.5   &   0.400  \\
0.50 &   10.2   &   0.752  \\
0.55 & \, 9.80  &   1.16   \\
0.60 & \, 9.33  &   1.63   \\
0.65 & \, 8.77  &   2.18   \\
0.70 & \, 8.10  &   2.82   \\
0.75 & \, 7.33  &   3.57   \\
0.80 & \, 6.43  &   4.47   \\
0.85 & \, 5.40  &   5.58   \\
0.90 & \, 4.20  &   7.01   \\
0.95 & \, 2.75  &   9.10   \\
1    & \, 0     &  \!\!\!\!\,14.7
\end{tabular}
\end{ruledtabular}
\end{table}
We note that the short-distance coefficient $d_1'(z)$ becomes negative
for some values of $z$. The complete fragmentation function, to all
orders in $v$, is an integral of the square of a quantity and is,
therefore, positive. However, the individual contributions in the $v$
expansion need not be positive. 

We estimate the fragmentation probability by integrating the
fragmentation function $D^H_1(z)$ over the longitudinal fraction $z$:
\begin{equation}
\int_0^1 dz \; D^H_1(z)= 8.29\times 10^{-4}\cdot \frac{\alpha_s^3}{m^3}
                     \langle\mathcal{O}_1^H\rangle
        \left[1 +2.45 \;v_1^2 +O(v^4)\right].
\label{integrated-color-singlet}
\end{equation}
Our result for the term of leading order in $v$ in
Eq.~(\ref{integrated-color-singlet}) is in agreement with the result of
Ref.~\cite{BY-gfrag-LO}.

\section{Discussion\label{sec:discuss}}

We have computed the contributions of leading order in $v$ and the
relativistic corrections of relative order $v^2$ to the fragmentation
function for a gluon to fragment into a ${}^3S_1$ heavy-quarkonium
state. We have computed both the contribution in which the produced
$Q\overline Q$ pair is in a color-octet state and the contribution in
which the $Q\overline Q$ pair is in a color-singlet state. Our results
of leading order in $v$ agree with those of
Refs.~\cite{Braaten:1994kd,BL:gfrag-NLO} for the color-octet
contribution and with those of Ref.~\cite{BY-gfrag-LO} for the
color-singlet contribution. Our results for the corrections of relative
order $v^2$ are new.

We estimate the relative sizes of the relativistic corrections for
fragmentation into $J/\psi$ by taking $v_1^2$ and $v_8^2$ from the 
Gremm-Kapustin relation \cite{Gremm:1997dq}
\begin{equation}
v_1^2=v_8^2=\frac{M-2m_{\rm pole}}{m_{\rm QCD}},
\label{gremm-kapustin}
\end{equation}
where $m_{\rm pole}$ is the pole mass and $m_{\rm QCD}$ is the mass that
appears in the NRQCD action. The Gremm-Kapustin relation follows from
the equations of motion of NRQCD and is accurate up to corrections of
relative order $v^2$. In the original work of Gremm and Kapustin
\cite{Gremm:1997dq}, a relation was given only for $v_1$. In
Eq.~(\ref{gremm-kapustin}), we have included the Gremm-Kapustin relation
for $v_8$, which can be derived in exactly the same manner as the
Gremm-Kapustin relation for $v_1$. Dimensional regularization of the
matrix elements $\langle\mathcal{P}_1^H\rangle$ and
$\langle\mathcal{P}_8^H\rangle$ is implicit in the Gremm-Kapustin
relation \cite{Bodwin:1998mn}. Taking $m_{\rm QCD}=m_{\rm pole}=1.4$~GeV
and $M_{J/\psi}=3.097$~GeV, we obtain $v_1^2=v_8^2=0.21$. One should
regard this as only a rough estimate of the sizes of $v_1$ and $v_8$. In
fact, for $m_{\rm pole}$ in the range $1.2\hbox{ GeV}<m_{\rm
pole}<1.6\hbox{ GeV}$, which corresponds to the latest Particle Data
Group compilation \cite{Hagiwara:fs}, the values of $v_1^2$ and $v_8^2$
given by Eq.~(\ref{gremm-kapustin}) can even become
negative.\footnote{Note that negative values of $v_1^2$ are allowed
since, owing to the subtractions of power divergences that are implicit
in dimensional regularization, the corresponding matrix element is not
positive definite.} On the other hand, the estimate for $v_1$ that we
obtain is in accordance with expectations from the NRQCD
velocity-scaling rules \cite{BBL}, and it lies in the central part of
the range $0.03<v_1^2<0.6$, which follows from a lattice calculation
\cite{bks-1} under the assumption that $m_c$ lies in the range
$1.2\hbox{ GeV}<m_c<1.6\hbox{ GeV}$. Inserting $v_1^2=v_8^2=0.21$ into
Eqs.~(\ref{color-octet-final}) and (\ref{integrated-color-singlet}), we
find that the relativistic corrections change the short-distance
coefficients for the color-octet contribution and the color-singlet
contribution by about $-40\%$ and $50\%$, respectively.

We see that, in the case of the $J/\psi$, the estimated relativistic
corrections are quite large in comparison with the contributions of
leading order in $v^2$. This is not unexpected, given that such large
relativistic corrections also appear in charmonium decays.\footnote{See,
for example, Ref.~\cite{Bodwin:2002hg}.} Nevertheless, these large
relativistic corrections cast some doubt on the validity of the $v$
expansion for charmonium. One can hope that, as is the case in
charmonium decays \cite{Bodwin:2002hg}, the corrections of relative
order $v^4$ will turn out to be significantly smaller than the
corrections of relative order $v^2$.

The value of the color-singlet matrix element $\langle
\mathcal{O}^{H}_1\rangle$ is fixed by quarkonium decay rates. Therefore,
the large relativistic correction to the short-distance coefficient of
the color-singlet contribution to the fragmentation function will
directly affect the theoretical predictions for quarkonium production
rates. However, in the case of $J/\psi$ production at the Tevatron, the
color-singlet fragmentation contribution is less than 5\% of the total
theoretical prediction over a wide range of $p_T$
\cite{Braaten:1994xb,Kniehl:1998qy,kramer}. Therefore, the relativistic
correction to the color-singlet short-distance coefficient will have
little effect on the predictions for either the $J/\psi$ production rate
or polarization at the Tevatron.

The color-octet matrix element $\langle \mathcal{O}^{H}_8\rangle$ in
Eq.~(\ref{color-octet-final}) is, at present, obtained by fitting the
Tevatron data for $J/\psi$ production to the complete theoretical
expression, which is dominated at the largest values of $p_T$ by the
color-octet fragmentation contribution. Therefore, the large decrease in
the short-distance coefficient for the color-octet contribution will
result in a corresponding large increase in the fitted value of the
matrix element $\langle \mathcal{O}^{H}_8\rangle$. The net result is
that the phenomenology of $J/\psi$ production at the Tevatron, for
either the $p_T$ distribution or the polarization, will be largely
unaffected by the relativistic correction to the color-octet
short-distance coefficient. Since, in leading order in $\alpha_s$, the
matrix element $\langle \mathcal{O}^{H}_8\rangle$ typically appears only
in the fragmentation contribution to a process, we expect the
phenomenology of other processes to be similarly unaffected. We note
that, even with the increase in the fitted value of $\langle
\mathcal{O}^{H}_8\rangle$ that would result from the relativistic
correction, the value of that matrix element would still be somewhat
smaller than is expected from the velocity scaling rules.\footnote{See,
for example, Ref.~\cite{kramer} for fitted values of the matrix
elements.} However, the relativistic correction may be important in
comparing phenomenological values of  $\langle \mathcal{O}^{H}_8\rangle$
with future lattice calculations.

\begin{acknowledgments}
We thank Eric Braaten for making a number of useful comments and,
particularly, for providing us with detailed information on the
phase-space-integration method of Ref.~\cite{BY-gfrag-LO}. We also thank
Andrea Petrelli for the use of some of his \textsc{mathematica} code.
Work in the High Energy Physics Division at Argonne National Laboratory
is supported by the U.~S.~Department of Energy, Division of High Energy
Physics, under Contract No.~W-31-109-ENG-38.
\end{acknowledgments}
\appendix
\section{Lorentz Transformation for the Fragmentation Function
\label{app:boost}}

We wish to construct a Lorentz transformation that changes the
transverse components of the hadron's momentum while leaving the $+$
components of all four-vectors unchanged. We will construct a particular
transformation from a frame in which the fragmenting gluon has vanishing
transverse momentum to a frame in which the quarkonium has vanishing
transverse momentum. Such a transformation was already discussed in
Ref.~\cite{CS}. Here, we give an explicit construction.

Given a basis set of four linearly independent four-vectors that span
the four-dimensional space-time, a transformation is a Lorentz
transformation iff it leaves all scalar products of the basis
four-vectors invariant. Using the notation $V=(V^+,V^-,V^1,V^2)$, we
choose for our basis set in the original Lorentz frame
\begin{subequations}
\label{orig-vecs}
\begin{eqnarray}
k&=&\left(k^+,\frac{k^2}{2k^+}
          ,0,0\right),\\
p&=&\left(zk^+,\frac{p^2+p^2_\perp}{2zk^+}
           ,p_\perp,0\right),\\
n&=&(0,1,0,0),\\
e_2&=&(0,0,0,1),
\end{eqnarray}
\end{subequations}
where we have taken the $1$ direction to be along $\bm{p}_\perp$. We
assume that $n$ is invariant under the Lorentz transformation, which
implies that the $+$ components of all four-vectors are also invariant.
We also assume, for simplicity, that $e_2$ is invariant. Then, the
transformed basis vectors are
\begin{subequations}
\label{transformed-vecs}
\begin{eqnarray}
k'&=&
\left(k^{+},\frac{k^2+p^2_\perp/z^2}{2k^+}
             ,\frac{-p_\perp}{z},0\right),\\ 
p'&=&\left(zk^{+},\frac{p^2}{2zk^+}
           ,0,0\right),\\
n'&=&(0,1,0,0),\\
e_2'&=&(0,0,0,1).
\end{eqnarray}
\end{subequations}
The transformed vectors are completely fixed by the requirement that
all of the scalar products be invariant under the transformation and
the requirements that $\bm{p}'_\perp=\bm{0}_\perp$, $n'=n$, and
$e_2'=e_2$.

Using Eqs.~(\ref{orig-vecs}) and (\ref{transformed-vecs}), we can
construct an explicit transformation matrix $L$ in the light-cone basis
$\bar n=(1,0,0,0)$, $n=(0,1,0,0)$, $e_1=(0,0,1,0)$, and $e_2=(0,0,0,1)$.
$L$ is defined by $V'=LV$, where $V$ is any four-vector, expressed as a
column vector. It follows that $L$ is given in terms of the
untransformed and transformed light-cone basis vectors by
\begin{eqnarray}
L=
\left(
\begin{array}{rrrr}
\bar n'\cdot n& 
n'\cdot n& 
e_1'\cdot n&
e_2'\cdot n \\[.3em]
\bar n'\cdot \bar n&
n'\cdot \bar n&
e_1'\cdot \bar n&
e_2'\cdot \bar n \\[.3em]
-\bar n'\cdot  e_1&    
-n'\cdot e_1&   
-e_1'\cdot e_1&
-e_2'\cdot e_1\\[.3em]
-\bar n'\cdot e_2&    
-n'\cdot e_2&   
-e_1'\cdot e_2&
-e_2'\cdot e_2\\
\end{array}
\right).
\label{eq:L0}
\end{eqnarray}
From Eq.~(\ref{orig-vecs}), we see that the light-cone basis vectors can be 
written in terms of $k$, $p$, $n$, and $e_2$ as 
\begin{subequations}
\label{lc-basis}
\begin{eqnarray}
\bar n&=&
\frac{1}{k^+}\left(k-n\;\frac{k^2}{2k^+}\right),
\\
n&=&n,\\
e_1&=&
\frac{1}{2p_\perp}\left[2(p-zk)
                       +n\;\frac{z^2k^2-p^2-p_\perp^2}{zk^+}
                       \right],\\
e_2&=&e_2.
\end{eqnarray}
\end{subequations}
Then, the transformed light-cone basis vectors are
\begin{subequations}
\label{lc-basis-p}
\begin{eqnarray}
\bar n'&=&
\frac{1}{k^+}\left(k'-n\;\frac{k^2}{2k^+}\right),
\\
n&=&n,\\
e_1'&=&
\frac{1}{2p_\perp}\left[2(p'-zk')
                       +n\;\frac{z^2k^2-p^2-p_\perp^2}{zk^+}
                       \right],\\
e_2'&=&e_2.
\end{eqnarray}
\end{subequations}
Substituting Eqs.~(\ref{lc-basis}) and (\ref{lc-basis-p}) into
Eq.~(\ref{eq:L0}) and using Eqs.~(\ref{orig-vecs}) and
(\ref{transformed-vecs}), we obtain
\begin{eqnarray}
L=
\left(
\begin{array}{cccc}
1&0&0&0\\
\frac{p^2_\perp}{2(p^+)^2}
&1&-\frac{p_\perp}{p^+}&0\\
-\frac{p_\perp}{p^+}&0&1&0\\
0&0&0&1
\end{array}
\right).
\label{eq:L2}
\end{eqnarray}
The inverse transformation from the primed coordinates to the unprimed 
coordinates is given by $L^{-1}=L(p_\perp\to -p_\perp)$.
\section{Phase Space for the color-octet contribution
\label{phase-space}}
In this appendix we describe the manipulations that we carry out on
the phase space for the color-singlet contribution to the fragmentation
function [Eq.~(\ref{eq:phase-space-2})] in order to put it into a form
that is suitable for numerical integration. We write the phase-space
integration as
\begin{equation}
\int_\Phi=
\int_{0}^{\infty} de_a\;
\int_{0}^{\infty} de_b\;
\int_{0}^{1-z}dy
\int_0^{2\pi} \frac{d\phi}{2\pi}
\;\Theta\;,
\end{equation}
where $\Theta=\theta\left(e_a-\frac{y}{2z}\right)%
\theta\left(e_b-\frac{w}{2z}\right)$. 

Since the integrand depends on $\phi$ only through the variable $x$
[Eq.~(\ref{x})], it is an even function of $\phi$. Therefore, we may
restrict the range of $\phi$ to $0$ to $\pi$ and double the integrand:
$\int_0^{2\pi} d\phi/(2\pi)\to \int_0^{\pi} d\phi/\pi$. In the range
$0\leq\phi\leq\pi$, $\phi$ is a single-valued function of $x$.
Consequently, we can make a change of integration variables in which we 
replace $\phi$ with $x$. Using Eq.~(\ref{ea-eb-x}), we rewrite $x$ as
\begin{subequations}
\label{eq:x-phi}
\begin{equation}
x=\frac{1}{2}\left(\alpha^2+\beta^2-2\alpha\beta\cos\phi \right)\;,
\label{eq:x-phi-a}
\end{equation}
where
\begin{eqnarray}
\alpha&=&\sqrt{\frac{2w}{z}\left(e_a-\frac{y}{2z}\right)}\;,
\\
\beta&=&\sqrt{\frac{2y}{z}\left(e_b-\frac{w}{2z}\right)}\;.
\end{eqnarray}
\end{subequations}
Note that, owing to the constraint $\Theta$, $\alpha$ and $\beta$ are
real and positive. The Jacobian $J$ for the change of the variables is
given by
\begin{subequations}
\label{eq:phi-x}
\begin{equation}
J=\left(\frac{dx}{d\phi}\right)^{-1}
=\frac{1}{\alpha\beta\sin\phi}=
\frac{z}{\sqrt{-A+2By-Cy^2}}=\frac{z}{\sqrt{C(y-y_-)(y_+-y)}},
\label{jacobian}
\end{equation}
where
\begin{eqnarray}
A&=&[zx-(1-z)e_a]^2,
\\
B&=&-zx\left(e_a-e_b+\frac{1-z}{z}\right)+(1-z)e_a(e_a+e_b),
\\
C&=&(e_a+e_b)^2-2x,
\\
y_\pm&=&\frac{B\pm\sqrt{D}}{C},
\\
D&=&B^2-AC=2z^2x(2e_ae_b-x)
\left[\frac{1-z}{z}\left(e_a+e_b-\frac{1-z}{2z}\right)-x\right].
\label{eq:D}
\end{eqnarray}
\end{subequations}
Using Eq.~(\ref{jacobian}), we arrive at the following form for the 
phase space:
\begin{eqnarray}
\int_\Phi&=&
\int_{0}^{\infty} de_a
\int_{0}^{\infty} de_b\int_0^{1-z} dy\int_0^{\pi}\frac{d\phi}{\pi}\;
\Theta\nonumber\\
&=& \int_{0}^{\infty} de_a\int_{0}^{\infty} de_b\int dx\int dy\;
\frac{z\Theta}{\pi\sqrt{C(y-y_-)(y_+-y)}}\;,
\label{eq:x-y}
\end{eqnarray}
where we have interchanged the $x$ and $y$ integrations, and we have not
yet specified the ranges of integration on the right-hand side of
Eq.~(\ref{eq:x-y}).

Now let us work out the ranges of integration. The denominator on the
right-hand side of Eq.~(\ref{eq:x-y}) has zeros at $y=y_+$ and $y=y_-$.
(There is no zero at $C=0$, since $y_+$ and $y_-$ become infinite at
that point.) As can be seen from Eq.~(\ref{jacobian}), the denominator
of Eq.~(\ref{eq:x-y}) is proportional to $\sin\phi$. Therefore, its
zeros correspond to $\phi=0$ and $\phi=\pi$, which are the end points of
the $\phi$ integration. We conclude that the range of $y$ is restricted
to $y_-\leq y \leq y_+$. In order for $y$ to have a nonzero range, we
must have $D>0$. This implies that $x$ lies either below both zeros of
$D(x)$ or above both zeros of $D(x)$. We now argue that the latter range
is unphysical. Clearly, for $x$ large enough, then Eq.~(\ref{eq:x-phi})
has solutions only for $y$ lying outside the physical region $0\leq
y\leq 1-z$ and $\Theta(y)=1$. Furthermore, $y$ is on the boundary of the
physical region iff $x$ is equal to a zero of $D(x)$. [To see this, note
that the zeros of $D(x)$ occur precisely when the range of $y$ vanishes.
From Eq.~(\ref{eq:x-phi-a}), we see that the range of $y$ vanishes iff
$\alpha=0$ or $\beta=0$. But $\alpha=0$ or $\beta=0 $ iff $y=0$ or
$y=1-z$ or $\Theta=0$, that is, iff $y$ is on the boundary of the
physical region.] Then, by continuity, we conclude that $y$ lies in the
physical region iff $x$ is restricted to be less than either of the zeros
of $D$. Therefore, we require that $x$ satisfy
\begin{equation}
x\leq x^{\textrm{max}}(z,e_a,e_b)=
\textrm{Min}\left[\;2e_ae_b,
\frac{1-z}{z}\left(e_a+e_b-\frac{1-z}{2z}\right)
              \right].
\end{equation}
Since, in this range of $x$, $y$ automatically satisfies
$0\leq y\leq 1-z$ and $\Theta(y)=1$, we can drop those
explicit constraints on $y$. For fixed $z$, the constraint $\Theta=1$
implies that $e_a+e_b\geq (1-z)/(2z)$. This constraint has already been
satisfied by virtue of the constraint $x\leq x^{\textrm{max}}$. However,
for purposes of improving numerical-integration efficiency, we can
explicitly impose it on the range of $e_b$. Taking all of the constraints
on the integration variables into account, we have for the phase-space
integration
\begin{equation}
\int_\Phi
=
\int_0^\infty de_a
\int_{\textrm{Max}\left[\;0,\frac{1-z}{2z}-e_a\right]}^\infty de_b
\int_0^{x^{\textrm{max}}}dx
\int_{y_-}^{y_+}dy\;\frac{z}{\pi\sqrt{C(y-y_-)(y_+-y)}}\;.
\end{equation}

It turns out that there is no $y$ dependence in the denominator of the
integrand and that the numerator of the integrand is a polynomial in $y$
of degree two. Therefore, the $y$ integration is easily performed:
\begin{equation}
Y_n\equiv
\int_{y_-}^{y_+}dy\;
\frac{y^n}{\pi\sqrt{(y-y_-)(y_+-y)}}
=
\sum_{r=0}^{r\le n/2}
\frac{n!\,
      \Gamma\left(r+\frac{1}{2}\right)
       \;B^{n-2r} \;(B^2-AC)^r }
     {r!\,(2r)!\,(n-2r)!\,\Gamma\left(\frac{1}{2}\right)\,
      C^n}\;.
\label{eq:y-integral}
\end{equation}
The explicit forms that are needed for the integrand in the 
present calculation are
\begin{subequations}
\label{eq:y-integral-table}
\begin{eqnarray}
Y_0&=&1,\\
Y_1&=&\frac{B}{C},\\
Y_2&=&\frac{3B^2-AC}{2C^2}.
\end{eqnarray}
\end{subequations}
In carrying out the integrations over $z$, $e_a$, $e_b$, and $x$
numerically, we made use of the adaptive Monte Carlo routine
\textsc{vegas} \cite{vegas}. We also checked our results using the
built-in numerical-integration package in \textsc{mathematica}
\cite{MATH}. Special care must be taken in carrying out the $x$
integration numerically. In the limits $e_a\to 0$ or  $e_b\to 0$, the
range of the $x$ integration vanishes. This can cause a large round-off
error in the Monte Carlo integration. We avoided this problem by making a
further change of variables $x=2e_ae_bt$. The integration interval for
$t$ is $0<t<x^{\textrm{max}}/(2e_ae_b)$.

\end{document}